\documentclass[preprint,aps,prd,showpacs,floatfix]{revtex4-1}
\usepackage{times}
\usepackage {euscript}
\usepackage{amsthm}
\usepackage{graphicx}
\usepackage{amsmath}
\usepackage{amssymb}
\usepackage{amsfonts}
\usepackage[english]{babel}
\usepackage{epstopdf}
\usepackage{multirow,makecell,array}
\usepackage{mathtools}
\usepackage{epstopdf}
\usepackage{color}
\newcommand{\bfr}{{\bf r}}

\newcommand{\bfp}{{\bf p}}
\newcommand{\bfA}{{\bf A}}
\newcommand{\bfD}{{\bf D}}
\newcommand{\balpha}{{\mbox{\boldmath$\alpha$}}}
\newcommand{\bnabla}{{\mbox{\boldmath$\nabla$}}}
\newcommand{\be}{\begin{eqnarray}}
\newcommand{\ee}{\end{eqnarray}}
\newcommand{\la}{\langle}
\newcommand{\ra}{\rangle}
\newcommand{\veps}{\varepsilon}

\newcommand{\ket}[1]{|#1\rangle}

\usepackage{ulem}

\begin{document}
\thispagestyle{empty}

    \title{Relativistic nuclear recoil effect on the $g$ factor of highly charged boronlike ions}

\author{I.~A.~Aleksandrov}
\author{D.~A.~Glazov}
\author{A.~V.~Malyshev}
\author{V.~M.~Shabaev}
\author{I.~I.~Tupitsyn}
\affiliation{Department of Physics, St.~Petersburg State University, 
Universitetskaya 7/9, 199034 St.~Petersburg, Russia
\vspace{10mm}
}

\begin{abstract}
The nuclear recoil effect on the $g$ factor of the ground state of highly charged B-like ions is studied to first order in the electron-to-nucleus mass ratio $m/M$. The leading one-electron and two-electron recoil contributions are calculated nonperturbatively in the parameter $\alpha Z$ within the rigorous QED formalism. The interelectronic-interaction correction to the nuclear recoil effect is evaluated in the Breit approximation by means of perturbation theory to the first order in $1/Z$. The higher-order interelectronic-interaction corrections are taken into account partially by performing the calculations with the effective screening potential. The results obtained represent the most accurate up-to-date treatment of the nuclear recoil effect on the $g$ factor of highly charged B-like ions in the range $Z = 20$--$92$.

\end{abstract}

\maketitle
\section{Introduction}\label{sec:intro}
The last two decades have seen a combination of intensive theoretical and experimental investigations of the $g$ factor of highly charged ions which delivered the stringent tests of bound-state quantum electrodynamics (QED) in the presence of external magnetic field (see, e.g., Refs.~\cite{sturm:17:a,sha15chem} and references therein). For instance, the comparison between the experimental data~\cite{haf00,ver04,stu11,stu13,stu14} and the theoretical predictions (see Ref.~\cite{zat17} and references therein) for H-like ions has yielded the most accurate value of the electron mass~\cite{stu14}. Investigations of Li-like ions have provided an access to the many-electron QED effects on the $g$ factor \cite{wag13,vol14a,yerokhin:17:pra-2}. Moreover, one may expect that the forthcoming experiments with highly charged ions will bring an independent determination of the fine-structure constant~\cite{sha06,vol14,yerokhin:2016:100801}. 

However, the impending studies aim not only at more accurate determination of fundamental constants but also at probing bound-state QED in a conceptually different regime where the nucleus can no longer be considered as a mere source of the strong Coulomb field. As the nucleus has a finite mass, one should take into account the corresponding recoil effect, which requires more sophisticated theoretical methods beyond the usual Furry picture be developed. In this regard, in the recent study~\cite{malyshev17}, it was shown that the nontrivial QED contribution to the nuclear recoil effect on the $g$ factor can be experimentally examined at strong-coupling regime in the so-called specific difference of the ground-state $g$-factor values of heavy H-like and Li-like ions of the same isotope. Recent measurement of the $g$-factor isotope shift in Li-like calcium \cite{koel16} has demonstrated the feasibility of the experimental investigations of this kind. Evaluation of the nuclear recoil effect within the most advanced methods available to date has improved the agreement between theory and experiment for Li-like calcium and provided the most accurate theoretical predictions for Li-like ions in the range $Z=10\text{--}92$ \cite{sha17l,shabaev:18:pra}.

The present study is focused on the nuclear recoil effect on the ground-state $g$ factor of B-like ions. In contrast to $s$ states of, e.g., H- and Li-like ions, for $p$ states this effect possesses a nonzero nonrelativistic limit and is generally more pronounced. Nevertheless, the theoretical uncertainty of the recoil contribution in the case of B-like ions is still much larger than that for H-like and Li-like systems due to the lack of relativistic calculations. The interest to these theoretical studies is also maintained by the $g$-factor measurements in B-like ions anticipated in the near future. Presently implemented ARTEMIS experiment at GSI~\cite{lin13,vogel:18:ap} and ALPHATRAP experiment at the Max-Planck-Institut f\"ur Kernphysik (MPIK) \cite{sturm:17:a} are expected to attain the accuracy of $10^{-9}$ and better for the $g$ factors of medium and heavy few-electron ions, including B-like ones. Recent calculations of the correlation and QED effects resulted in the theoretical predictions with the accuracy up to $10^{-6}$ \cite{soria_orts07,glazov:13:ps,verdebout:14:adndt,shc15,marques:16:pra,agababaev:18:jpcs}, which demands detailed study of the nuclear recoil effect as well. This motivates us to carry out systematic QED calculations in order to obtain the most accurate up to date values of the nuclear recoil contribution to the ground-state $g$ factor of B-like ions.

In our previous study~\cite{glazov:18:os}, we evaluated this contribution to first order in the electron-to-nucleus mass ratio $m/M$ within the lowest-order relativistic (Breit) approximation for light B-like ions ($Z=10$--$20$) taking into account the first-order corrections due to the electron-electron interaction effects. In the present investigation, we extend our studies to B-like ions in the range $Z = 20$--$92$ and carry out the first evaluation of the nontrivial QED part of the recoil effect to all orders in $\alpha Z$. The influence of the correlation effects is also treated within the Breit approximation to the first order of perturbation theory. The higher-order interelectronic-interaction corrections are taken into account approximately by means of the effective potential. Relativistic units ($\hbar = 1$, $c = 1$) and Heaviside charge unit [$\alpha=e^2/(4\pi)$, $e<0$] are employed throughout the paper.
%
\section{Theoretical methods}
\label{sec:formulas}
Our theoretical description is based on the independent-electron approximation with the Dirac Hamiltonian for the spherically symmetric binding potential $V(r)$,
\begin{equation}
\label{eq:hD}
  h_\text{D} = -i \balpha \cdot \nabla + \beta m + V (r)
\,.
\end{equation}
We assume the external magnetic field to be directed along the $z$ axis, $\boldsymbol{\mathcal{H}}=\mathcal{H}{\bf e}_z$, and describe it by the classical vector potential of the form $\bfA_\text{cl} (\bfr) = [\boldsymbol{\mathcal{H}} \times \bfr ] /2$ and the corresponding contribution $-e\balpha\cdot\bfA_\text{cl}(\bfr) = \mu_0\mathcal{H}{m}\,[ \bfr \times \balpha ]_z$ to the Dirac Hamiltonian, where $\mu_0=-e/2m$ is the Bohr magneton. According to Refs.~\cite{sha01a,yelkhovsky:01,sha03}, in the case of one electron in the $\ket{a}$ state over closed shells, the nuclear recoil correction to the $g$ factor can be evaluated to zeroth order in $1/Z$ as
\begin{equation}
\Delta g = \frac{1}{\mu_0 m_a}\frac{1}{M}\frac{i}{2\pi}
\int \limits_{-\infty}^{\infty} d\omega\;
\Biggl[\frac{\partial}{\partial {\mathcal H}}
\la \tilde{a}|[p^k-D^k(\omega)+eA_\text{cl}^k]\tilde{G}(\omega+\tilde{\veps}_a)
[p^k-D^k(\omega)+eA_\text{cl}^k]
|\tilde{a}\ra
\Biggr]_{{\mathcal H}=0}\, ,
\label{06recoilt}
\end{equation}
where $m_a$ is the angular momentum projection of the $a$ state, $p^k=-i\nabla^k$ is the momentum operator, $D^k(\omega)=-4\pi\alpha Z\alpha^l D^{lk}(\omega)$, and
\begin{equation}\label{06photon}
D^{lk}(\omega,{\bf r}) = -\frac{1}{4\pi}\Bigl\{\frac
{\exp{(i|\omega|r)}}{r}\delta_{lk} +\nabla^{l}\nabla^{k}
\frac{(\exp{(i|\omega|r)}
-1)}{\omega^{2}r}\Bigr\}\,
\end{equation}
is the transverse part of the photon propagator in the Coulomb gauge. We also imply here and below the summation over the repeated indices. The tilde sign indicates that the corresponding quantities (the wave function~$|\tilde{a}\ra$, the energy~$\tilde{\veps}_a$, and the Dirac-Coulomb Green's function $\tilde{G}(\omega)$) are to be determined in the presence of magnetic field. The Dirac-Coulomb Green's function reads $\tilde{G}(\omega)=\sum_{\tilde{n}}|\tilde{n}\ra \la \tilde{n}|[\omega-\tilde{\veps}_n +  i\eta(\tilde{\veps}_n - \tilde{\veps}_\text{F})]^{-1}$, where $ \tilde{\veps}_\text{F} $ is the Fermi energy and $\eta \rightarrow 0$. As noted in Ref.~\cite{sha01a}, one can partially take into account the nuclear size correction to the recoil effect on the $g$ factor by using the extended-nucleus potential $V_{\rm nucl}(r)$ as $V(r)$ in $h_\text{D}$. Moreover, one can replace $V_\text{nucl}(r)$ with the effective potential $V_\text{eff}(r)=V_\text{nucl}(r)+V_\text{scr}(r)$ which includes some local screening potential $V_{\rm scr}(r)$. This allows one to partially take into account the interelectronic-interaction effects.

Expression (\ref{06recoilt}) incorporates both one- and two-electron nuclear recoil contributions. The one-electron part $\Delta g^\text{(1el)}$ is represented as a sum of the low-order and higher-order terms, $\Delta g^\text{(1el)} = \Delta g_\text{L}^\text{(1el)}+\Delta g_\text{H}^\text{(1el)}$. The former completely includes the contributions of orders $(\alpha Z)^0$ and $(\alpha Z)^2$ and can be derived from the relativistic Breit equation,
\begin{align}
\label{low}
  \Delta g_{\rm L}^\text{(1el)} =& \frac{1}{m_a} \frac{{m}}{M} 
    \la \delta a| \left[ \bfp^2 -\frac{\alpha Z}{r} \left( \balpha + \frac{(\balpha\cdot\bfr)\bfr}{r^2} \right) \cdot \bfp \right] | a \ra
\nonumber\\
  & -\frac{1}{m_a} \frac{m}{M} \la a | \left( [\bfr \times \bfp]_z - \frac{\alpha Z}{2r}[\bfr \times \balpha ]_z \right) | a \ra 
\,,
\end{align}
where $|\delta a\ra = \sum_n^{\veps_n\ne \veps_a} | n \ra \la n | [\bfr\times\balpha]_z | a \ra (\veps_a-\veps_n)^{-1}$. In the case of a point-like nucleus, this contribution can be evaluated analytically~\cite{sha01a}:
\begin{equation}
\label{breit_point}
  \Delta g_\text{L}^\text{(1el)}{\rm[p.n.]} = -\frac{m}{M} \, \frac{2\kappa^2 \varepsilon^2_a + \kappa m \varepsilon_a - m^2}{2m^2 j (j+1)} \,,
\end{equation}
where $\kappa$ is the relativistic angular quantum number of the $a$ state, $j$ is its total angular momentum, and $\varepsilon_a$ is the Dirac energy with the rest mass included. To leading orders in $\alpha Z$, for $2p_{1/2}$ state it yields
\begin{equation}
\Delta g_\text{L}^\text{(1el)} {\rm[p.n.]} =-\frac{m}{M}\left[ \frac{4}{3} - \frac{5}{12}(\alpha Z)^2  + \ldots \right] 
\,.
\end{equation}
In the case of a general binding potential, e.g., extended-nucleus potential $V_{\rm nucl}(r)$ or the effective screening potential $V_{\rm eff}(r)$, the low-order term is to be evaluated numerically according to Eq.~(\ref{low}).

The higher-order contribution $\Delta g_\text{H}^\text{(1el)}$ to the one-electron part can only be derived within the rigorous QED approach beyond the Breit approximation. This contribution, referred to as the one-electron QED recoil term, reads~\cite{sha01a}
\begin{align}
\label{QED_1el}
  \Delta g_\text{H}^\text{(1el)} =& \frac{1}{m_a} \frac{m}{M} \frac{i}{2\pi}
    \int \limits_{-\infty}^{\infty} d\omega\;
      \Bigl\{ \la \delta a|B^k_{-}(\omega) G(\omega+\veps_a)B^k_{+}(\omega)|a\ra + \la a|B^k_{-}(\omega) G(\omega+\veps_a)B^k_{+}(\omega) |\delta a \ra
\nonumber\\ 
  & + \la a|B^k_{-}(\omega) G(\omega+\veps_a) \left([\bfr\times\balpha]_z - \langle a | [\bfr\times\balpha]_z | a \rangle \right) G(\omega+\veps_a) B^k_{+}(\omega)|a\ra \Bigr\}
\,, 
\end{align}
where $B^k_{\pm}(\omega)=D^k(\omega)\pm [p^k,V]/(\omega+i0)$, $V(r)$ is either the potential of the nucleus $V_{\rm nucl}(r)$ or the effective potential $V_{\rm eff}(r)$ (see the discussion above), $G(\omega)=\sum_n|n\ra \la n|[\omega-\veps_n(1-i0)]^{-1}$ is the conventional Dirac-Coulomb Green's function.

The two-electron part of Eq.~(\ref{06recoilt}) valid to all orders in $\alpha Z$ reads as follows \cite{shc15}:
\begin{align}
  \Delta g^\text{(2el)} =
    & \frac{1}{m_a} \frac{m}{M} \sum_c \, \Biggl\{ 
      \epsilon_{3kl} \left( \la a |r^k| c \ra \la c |[p^l-D^l(\Delta)]| a \ra 
                          + \la a |[p^l-D^l(\Delta)]| c \ra \la c |r^k| a \ra \right)
\nonumber\\
  & - \la \delta a |[p^k-D^k(\Delta)]| c \ra \la c |[p^k-D^k(\Delta)]| a \ra
    - \la a |[p^k-D^k(\Delta)]| \delta c \ra \la c |[p^k-D^k(\Delta)]| a \ra
\nonumber\\
  & - \la a |[p^k-D^k(\Delta)]| c \ra \la \delta c |[p^k-D^k(\Delta)]| a \ra
    - \la a |[p^k-D^k(\Delta)]| c \ra \la c |[p^k-D^k(\Delta)]| \delta a \ra
\nonumber\\
  & + \left(
      \la a |\frac{D^k(\omega)}{d\omega}\Bigr|_{\omega=\Delta}| c \ra \la c |[p^k-D^k(\Delta)]| a \ra 
    + \la a |[p^k-D^k(\Delta)]| c \ra \la c |\frac{D^k(\omega)}{d\omega}\Bigr|_{\omega=\Delta}| a \ra 
    \right)
\nonumber\\
  & \times \left( 
      \langle a | [\bfr\times\balpha]_z | a \rangle - \langle c | [\bfr\times\balpha]_z | c \rangle 
    \right)
  \Biggr\}
\,, 
\label{QED_2el}
\end{align}
where $\Delta=\varepsilon_a-\varepsilon_c$, $\epsilon_{ikl}$ is the Levi-Civita symbol, $|\delta c\ra=\sum_n^{\veps_n\ne \veps_c}|n\ra\la n|[\bfr\times\balpha]_z|c\ra (\veps_c-\veps_n)^{-1}$, and the summation runs over all of the $1s$ and $2s$ closed-shell electronic states $|c\rangle$. It should be noted that in Li-like ions this contribution vanishes completely~\cite{sha17l,shabaev:18:pra}. In what follows, we will also represent this contribution as a sum of the low-order (Breit) and the higher-order parts, $\Delta g^\text{(2el)}=\Delta g_\text{L}^\text{(2el)}+\Delta g_\text{H}^\text{(2el)}$. The low-order part is obtained from Eq.~(\ref{QED_2el}) by setting $\Delta=0$ (the derivative of $D$ vanishes in this case) and dropping out the $\bfD\cdot \bfD$ products. The higher-order part $\Delta g_\text{H}^\text{(2el)}$ is the remainder.

In order to take into account the interelectronic-interaction effect on the nuclear recoil correction to the $g$ factor within the Breit approximation, it is convenient to represent the low-order parts, $\Delta g_\text{L}^\text{(1el)}$ and $\Delta g_\text{L}^\text{(2el)}$, as the contributions of the following operators~\cite{sha17l}:
\be 
\label{br2}
H_M^\text{magn}&=&-\mu_0 {\mathcal H}
  \frac{m}{M}
  \sum_{i, j}\Bigl\{[\bfr_i\times \bfp_j] - \frac{\alpha Z}{2r_j}\Bigl[\bfr_i\times\Bigl(\balpha_j
   +\frac{(\balpha_j\cdot\bfr_j)\bfr_j}{r_j^2}\Bigr)\Bigr]
  \Bigr\}
\,,\\
\label{br1}
H_M&=&\frac{1}{2M}\sum_{i,j}\Bigl[\bfp_i\cdot \bfp_j
  -\frac{\alpha Z}{r_i}\Bigr(\balpha_i+\frac{(\balpha_i\cdot\bfr_i)\bfr_i}{r_i^2}\Bigr)
  \cdot\bfp_j\Bigr]
\,.
\ee
The contribution of $H_M^\text{magn}$ is given by the average value,
\be 
\Delta g^\text{(0)}_\text{magn}&=&\frac{1}{\mu_0 {\mathcal H} m_a} \la A | H_M^\text{magn} | A \ra \, . 
\label{g_br2}
\ee
Here $|A\rangle$ is the many-electron wave function of the state of interest. This part contains the nonrelativistic limit of the nuclear recoil effect derived by Phillips~\cite{phi49}. Therefore, it dominates for low and middle values of $Z$ except for the $s$ states where this limit yields zero. The operator~(\ref{br1}) describes the nuclear recoil effect on the binding energies in the absence of external field \cite{sha98}. The corresponding contribution to the $g$ factor is given by the second-order term of the perturbation theory,
\begin{equation}
\label{g_br1}
  \Delta g^\text{(0)}_\text{nonmagn} = \frac{2}{\mu_0 {\mathcal H} m_a}\,{\sum_{N}}'\,\frac{\la A | H_M | N \ra \la N | H_\text{magn} | A \ra} {E_A - E_N}
\,,
\end{equation}
where the operator $H_\text{magn} = \mu_0\mathcal{H} m \sum_{j} [ \bfr_j \times \balpha_j ]_z$ describes the interaction with the magnetic field. The summation in Eq.~(\ref{g_br1}) runs over the complete spectrum of the many-electron states $|N\rangle$ constructed from the Dirac wave functions, including the one-electron negative-energy excitations. The prime here and below indicates that the terms with $E_N=E_A$ are excluded from the summation. The superscripts (0) indicate that these terms are of zeroth order with respect to the interelectronic interaction. In the case of one electron over the closed shells, e.g., for B-like ions, Eqs.~(\ref{g_br2}) and (\ref{g_br1}) together yield the same result as Eq.~(\ref{low}) and the low-order part of Eq.~(\ref{QED_2el}),
\begin{equation}
  \Delta g_\text{L}^\text{(1el)}+\Delta g_\text{L}^\text{(2el)} =
    \Delta g^\text{(0)}_\text{nonmagn} + \Delta g^\text{(0)}_\text{magn}
\,.
\end{equation}

The influence of the electron-electron interaction on the nuclear recoil contribution to the $g$ factor is considered with the use of the Dirac-Coulomb-Breit Hamiltonian,
\begin{equation}
\label{eq:HDCB}
  H_\text{DCB} = \Lambda_+ \left( \sum_j h_\text{D}(j) + H_\text{magn} + H_\text{int} \right) \Lambda_+
\,,
\end{equation}
where the Coulomb-Breit operator $H_\text{int}$ represents the interelectronic interaction, 
\begin{equation}
\label{int-int}
  H_\text{int} = \alpha \sum_{i<j} \Biggl [ \frac{1}{r_{ij}} - \frac{ {\balpha}_i \cdot {\balpha}_j }{ r_{ij} }
    - \frac{1}{2} ( {\balpha}_i \cdot {\bnabla}_i ) ( {\balpha}_j \cdot {\bnabla}_j ) r_{ij} \Biggr ]
\,,
\end{equation}
and $\Lambda_+$ is the positive-energy-states projection operator, constructed as the product of the one-electron projectors. If the Dirac Hamiltonian $h_\text{D}$ includes the screening potential, the corresponding counterterm $\delta H_\text{int}=-\sum_j V_{\rm scr}(r_j)$ has to be included in $H_\text{int}$.
To the first order of the perturbation theory in $H_\text{int}$, the magnetic part (\ref{g_br2}) of the nuclear recoil contribution to the $g$ factor acquires the following correction:
\begin{equation} 
\label{g_Z_magn}
  \Delta g^\text{(1)}_\text{magn} = \frac{2}{\mu_0 {\mathcal H} m_a}\,{\sum^{(+)}_N}{\vphantom{\sum}}'\,\frac{\la A | H^\text{magn}_M | N \ra \la N | H_\text{int} | A \ra} {E_A - E_N}
\,,
\end{equation}
where the plus sign over the sum indicates that the intermediate $|N\rangle$ states are constructed only from the positive-energy one-electron Dirac wave functions. This correction completely takes into account the first-order term in $1/Z$ and can partially incorporate the higher-order terms due to employing the effective screening potential. The first-order correction to the nonmagnetic part (\ref{g_br1}) can be written in the following form
\begin{equation}
\label{eq:g_Z_nonmagn}
  \Delta g^\text{(1)}_\text{nonmagn} = 
    \frac{2}{\mu_0 m_a}\,\frac{\partial}{\partial {\mathcal H}}
    \left.
       {\sum^{(+)}_N}{\vphantom{\sum}}'\,\frac{\la \tilde A | H_M | \tilde N \ra \la \tilde N | H_\text{int} | \tilde A \ra} {\tilde E_A - \tilde E_N}
    \right|_{\mathcal H=0}
\,,
\end{equation}
where the tilde sign indicates that the wave functions $\ket{\tilde A}$, $\ket{\tilde N}$ and the energies $\tilde E_A$, $\tilde E_N$ are evaluated in the presence of the magnetic field $\mathcal H$, similarly to Eq.~(\ref{06recoilt}). In this case, the positive-energy-states projector $\Lambda_+$ also corresponds to the Dirac equation in the presence of magnetic field. Derivatives of $\Lambda_+$ with respect to the magnetic field give rise to the necessary contribution of the negative energy states \cite{glazov:04:pra,tupitsyn:05:pra}. Equation~(\ref{eq:g_Z_nonmagn}) can be rewritten in the explicit form involving the matrix elements of $H_\text{magn}$. However, since this form is rather cumbersome, it is not presented here.

\section{Results and discussion}\label{sec:results}
%
%
In this section, we present the results of our evaluation of the contributions to the nuclear recoil effect according to Eqs.~(\ref{low}), (\ref{QED_1el}), (\ref{QED_2el}), (\ref{g_Z_magn}), and (\ref{eq:g_Z_nonmagn}). We perform the necessary angular integrations analytically and calculate the radial parts and sum over the intermediate states numerically. To this end, we use the finite-basis sets constructed from $B$ splines~\cite{Johnson:1988:307,sap96} within the dual-kinetic-balance approach~\cite{sha04}. The Fermi model is employed to describe the nuclear charge distribution, and the nuclear charge radii are taken from Ref.~\cite{ang13}. Performing the calculations with an effective potential allows one to take into account the higher-order interelectronic-interaction effects only partly. In order to have a reliable estimation of the higher-order contributions, we carry out the evaluation of the nuclear recoil effect on the $g$ factor of B-like ions with several different types of the effective potential. In particular, we employ the core-Hartree (CH), Perdew-Zunger (PZ)~\cite{per81}, Kohn-Sham (KS)~\cite{pot:KS}, and local Dirac-Fock (LDF)~\cite{sha05} potentials (see also, e.g., Ref.~\cite{sap02}). 

The low-order contributions to the $g$ factor of B-like ions to zeroth order in $1/Z$ are conveniently represented via the function $A(\alpha Z)$,
\begin{equation}
\label{A_def}
  \Delta g^{(0)} = \frac{m}{M}  A(\alpha Z)  
\,.
\end{equation}
%
%
In Table~\ref{table:1el_Breit} we display the one-electron low-order part $\Delta g_\text{L}^\text{(1el)}$ in terms of the function $A (\alpha Z)$. While the results of the calculations with the pure Coulomb potential induced by the extended nucleus are shown in the first column, the other columns present the data obtained with four types of the effective screening potential.
\begin{table}
\centering
\setlength{\tabcolsep}{1.2em}
\caption{The low-order one-electron recoil contribution $\Delta g_\text{L}^\text{(1el)}$ to the $g$ factor of the $2p_{1/2}$ state. The results are presented in terms of the function $A(\alpha Z)$ defined by Eq.~(\ref{A_def}). The indices Coul, CH, PZ, KS, and LDF refer to the calculations with the Coulomb and various screening potentials (see the text). 
\label{table:1el_Breit}}
\begin{tabular}{cr@{.}lr@{.}lr@{.}lr@{.}lr@{.}l}
\hline \hline
$Z$& \multicolumn{2}{c}{$A^{\rm(1el)}_{\rm Coul}$}  & 
     \multicolumn{2}{c}{$A^{\rm(1el)}_{\rm CH}$}       &
     \multicolumn{2}{c}{$A^{\rm(1el)}_{\rm PZ}$}       &
     \multicolumn{2}{c}{$A^{\rm(1el)}_{\rm KS}$}      &
     \multicolumn{2}{c}{$A^{\rm(1el)}_{\rm LDF}$}       \\
\hline
   20  &   $-$1&32441     &   $-$1&32556  &    $-$1&32537  &    $-$1&32538  &   $-$1&32548   \\
   30  &   $-$1&31311     &   $-$1&31486   &    $-$1&31456 &    $-$1&31458  &   $-$1&31473   \\
   40  &   $-$1&29700     &   $-$1&29939   &    $-$1&29897 &    $-$1&29899  &   $-$1&29920    \\
   50  &   $-$1&27579     &   $-$1&27885   &    $-$1&27829 &    $-$1&27831  &   $-$1&27861    \\
   60  &   $-$1&24901     &   $-$1&25282   &    $-$1&25209 &    $-$1&25212  &   $-$1&25252   \\
   70  &   $-$1&21607     &   $-$1&22071   &    $-$1&21978 &    $-$1&21981  &   $-$1&22034    \\
   80  &   $-$1&17619     &   $-$1&18177   &    $-$1&18059 &    $-$1&18063  &   $-$1&18133    \\
   82  &   $-$1&16731     &   $-$1&17309   &    $-$1&17186 &    $-$1&17189  &   $-$1&17263    \\
   90  &   $-$1&12862     &   $-$1&13527   &    $-$1&13378 &    $-$1&13381  &   $-$1&13473    \\
   92  &   $-$1&11817     &   $-$1&12505   &    $-$1&12349 &    $-$1&12352  &   $-$1&12448    \\
\hline \hline            
\end{tabular}
\end{table}

The two-electron nuclear recoil contribution to the $g$ factor of B-like ions in the zeroth order in $1/Z$ is presented in Table~\ref{table:2el} in terms of the function $A(\alpha Z)$ defined by Eq.~(\ref{A_def}). For each $Z$ the low-order part $\Delta g_{\rm L}^{\rm(2el)}$ of the two-electron contribution is given in the first line, whereas the higher-order correction $\Delta g_{\rm H}^{\rm(2el)}$ is shown in the second line. It is seen that the higher orders in $\alpha Z$ become considerably important for large values of the nuclear charge $Z$. While for $Z=20$ the higher-order part amounts to about $0.01\%$ of the total two-electron contribution, it reaches $11\%$ at $Z=92$.
\begin{table}
\centering
\setlength{\tabcolsep}{1.em}
\caption{The two-electron nuclear recoil contribution $\Delta g^\text{(2el)}$ to the ground-state $g$ factor of B-like ions in zeroth order in $1/Z$. The low-order part $\Delta g_\text{L}^\text{(2el)}$ and the higher-order part $\Delta g_\text{H}^\text{(2el)}$ are given separately in the lines labeled with L and H, respectively. The results are expressed in terms of the function $A(\alpha Z)$ defined by Eq.~(\ref{A_def}). 
\label{table:2el}}
\def\arraystretch{0.9}
\begin{tabular}{ccr@{.}lr@{.}lr@{.}lr@{.}lr@{.}l}
\hline \hline
$Z$& Part 
& \multicolumn{2}{c}{$A^{\rm (2el)}_{\rm Coul}$}
& \multicolumn{2}{c}{$A^{\rm (2el)}_{\rm CH}$}
& \multicolumn{2}{c}{$A^{\rm (2el)}_{\rm PZ}$} 
& \multicolumn{2}{c}{$A^{\rm (2el)}_{\rm KS}$} 
& \multicolumn{2}{c}{$A^{\rm (2el)}_{\rm LDF}$}
\\
\hline
   \multirow{2}{*}{20}  &   L & 0&54845     &   0&59158 &    0&59278  &    0&59869  &   0&59315    \\
     & H & $-$0&00008     &   $-$0&00006   &    $-$0&00006  &   $-$0&00006   &    $-$0&00006 \\
   \hline
   \multirow{2}{*}{30}  &   L & 0&54046     &   0&56825  &    0&56884 &    0&57225  &   0&56898    \\
    &   H   & $-$0&00040     &   $-$0&00033   &    $-$0&00034 &    $-$0&00033  &   $-$0&00034    \\
   \hline
   \multirow{2}{*}{40}  &   L & 0&52948     &   0&55063  &    0&55082  &    0&55321  &   0&55095   \\
     &   H   & $-$0&00127     &   $-$0&00111   &    $-$0&00113 &    $-$0&00112  &   $-$0&00112   \\
   \hline
   \multirow{2}{*}{50}  &   L & 0&51582     &   0&53339  &    0&53329  &    0&53513  &   0&53347    \\
     &   H   & $-$0&00313     &   $-$0&00282   &    $-$0&00287 &    $-$0&00284  &   $-$0&00285   \\
   \hline
   \multirow{2}{*}{60}  &   L & 0&49997     &   0&51539  &    0&51504 &    0&51654  &   0&51531    \\
     &   H   & $-$0&00663     &   $-$0&00610   &    $-$0&00618 &    $-$0&00614  &   $-$0&00614    \\
   \hline
   \multirow{2}{*}{70}  &   L & 0&48280     &   0&49682  &    0&49627 &    0&49754  &   0&49664    \\
     &   H   & $-$0&01273     &   $-$0&01189   &    $-$0&01203 &    $-$0&01198  &   $-$0&01196   \\
   \hline
   \multirow{2}{*}{80}  &   L & 0&46595     &   0&47896  &    0&47825  &    0&47937  &   0&47871   \\
     &   H   & $-$0&02311     &   $-$0&02182   &    $-$0&02208 &    $-$0&02202  &   $-$0&02193   \\
   \hline
   \multirow{2}{*}{82}  &   L & 0&46285     &   0&47568  &    0&47495 &    0&47605  &   0&47543    \\
     &   H   & $-$0&02594     &   $-$0&02454   &    $-$0&02482&    $-$0&02477  &   $-$0&02466    \\
   \hline
   \multirow{2}{*}{90}  &   L & 0&45252     &   0&46467  &    0&46389 &    0&46491  &   0&46440   \\
     &   H   & $-$0&04097     &   $-$0&03901  &    $-$0&03947 &    $-$0&03943  &   $-$0&03919    \\
   \hline
   \multirow{2}{*}{92}  &   L & 0&45072     &   0&46269  &    0&46192  &    0&46293  &   0&46242   \\
     &   H   & $-$0&04594     &   $-$0&04380   &    $-$0&04432 &    $-$0&04429  &   $-$0&04401    \\
\hline \hline            
\end{tabular}
\end{table}

The higher-order one-electron contribution is expressed in terms of the function $P(\alpha Z)$ defined by
\begin{equation} \label{P_def}
\Delta g^\text{(1el)}_\text{H} = \frac{m}{M} \frac{(\alpha Z)^3}{8} P(\alpha Z)
\,.
\end{equation}
We note that for $s$ states this contribution possesses the $(\alpha Z)^5$ behavior \cite{sha02b,sha17l} in contrast to the general $(\alpha Z)^3$ behavior for states with $l\neq 0$. The numerical results for the function $P(\alpha Z)$ for $2p_{1/2}$ state are presented in Table~\ref{table:1el_QED} for the Coulomb and four different effective screening potentials. It is seen that the QED contribution $\Delta g^\text{(1el)}_\text{H}$ to the nuclear recoil effect strongly increases with $Z$. 
\begin{table}
\centering
\setlength{\tabcolsep}{1.2em}
\caption{The higher-order (QED) one-electron recoil contribution $\Delta g_\text{H}^\text{(1el)}$ to the $g$ factor of the $2p_{1/2}$ state. The results are expressed in terms of the function $P(\alpha Z)$ defined by Eq.~(\ref{P_def}). \label{table:1el_QED}}
\begin{tabular}{cr@{.}lr@{.}lr@{.}lr@{.}lr@{.}l}
\hline \hline
$Z$
& \multicolumn{2}{c}{$P_{\rm Coul}$}
& \multicolumn{2}{c}{$P_{\rm CH}$}
& \multicolumn{2}{c}{$P_{\rm PZ}$} 
& \multicolumn{2}{c}{$P_{\rm KS}$} 
& \multicolumn{2}{c}{$P_{\rm LDF}$}
\\
\hline
   20  &   0&56851     &   0&45781   &    0&47084 &    0&45875  &   0&46551    \\
   30  &   0&65004     &   0&56209   &    0&57277 &    0&56352  &   0&56827    \\
   40  &   0&73423     &   0&65719   &    0&66676 &    0&65895  &   0&66255    \\
   50  &   0&82350     &   0&75251   &    0&76165 &    0&75482  &   0&75743    \\
   60  &   0&92423     &   0&85622   &    0&86551 &    0&85947  &   0&86098    \\
   70  &   1&05012     &   0&98201   &    0&99223 &    0&98689  &   0&98694    \\
   80  &   1&23001     &   1&15730   &    1&16973 &    1&16505  &   1&16289    \\
   82  &   1&27705     &   1&20262   &    1&21573 &    1&21119  &   1&20843    \\
   90  &   1&52714     &   1&44169  &    1&45885 &    1&45487  &   1&44887    \\
   92  &   1&61106     &   1&52148   &    1&54010 &    1&53628  &   1&52917   \\
\hline \hline            
\end{tabular}
\end{table}

Next, we turn to the first-order (in $1/Z$) interelectronic-interaction correction to the nuclear recoil effect on the $g$ factor, evaluated as a sum of Eqs.~(\ref{g_Z_magn}) and (\ref{eq:g_Z_nonmagn}),
\be 
\label{eq:low_part}
  \Delta g_{\rm L}^{(1)}=\Delta g^{(1)}_{\rm magn}+\Delta g^{(1)}_{\rm nonmagn} 
\,.
\ee
The results are presented in Table~\ref{table:1_Z_Breit} in terms of the function $B(\alpha Z)$,
%
\begin{align}
\label{B_def}
\Delta g^{(1)}_\text{L} &= \frac{m}{M} \frac{B(\alpha Z)}{Z}  
\,.
\end{align}
Despite a substantial discrepancy among the individual terms computed with different screening potentials, the complete Breit-approximation value, which is a sum of the low-order contributions from Tables~\ref{table:1el_Breit}, \ref{table:2el}, and \ref{table:1_Z_Breit}, turns out to be much more stable. This statement is demonstrated for $Z=20$ and $Z=92$ in Table~\ref{table:Breit_total}, where the zeroth-order term $A=A^\text{(1el)}+A^\text{(2el)}$, the first-order term $B/Z$, and their sum are presented for all the potentials considered. The corresponding KS and LDF values from Ref.~\cite{glazov:18:os} are presented as well. One observes a significantly better agreement among the total values than among the individual terms. A small difference between the present KS and LDF results and those from Ref.~\cite{glazov:18:os} is due to a modification of the procedures for designing the corresponding potentials in the present study.
\begin{table}
\centering
\setlength{\tabcolsep}{1.2em}
\caption{The $1/Z$ interelectronic-interaction correction $\Delta g^{(1)}_L$ to the nuclear recoil effect on the ground-state $g$ factor of B-like ions evaluated within the Breit approximation. The results are expressed in terms of the function $B(\alpha Z)$ defined by Eq.~(\ref{B_def}). 
\label{table:1_Z_Breit}}
\begin{tabular}{cr@{.}lr@{.}lr@{.}lr@{.}lr@{.}l}
\hline \hline
$Z$
& \multicolumn{2}{c}{$B_{\rm Coul}$}
& \multicolumn{2}{c}{$B_{\rm CH}$}
& \multicolumn{2}{c}{$B_{\rm PZ}$} 
& \multicolumn{2}{c}{$B_{\rm KS}$} 
& \multicolumn{2}{c}{$B_{\rm LDF}$}
\\
\hline
   20  &   1&8470     &   1&2024   &    1&1620 &    1&0305  &   1&1603    \\
   30  &   1&8300     &   1&1766    &    1&1414 &    1&0331  &   1&1451   \\
   40  &   1&8038     &   1&1498   &    1&1190  &    1&0202  &   1&1258   \\
   50  &   1&7661     &   1&1191     &    1&0914 &    0&9979  &   1&1004   \\
   60  &   1&7126     &   1&0831   &    1&0561 &    0&9660  &   1&0676    \\
   70  &   1&6377     &   1&0413   &    1&0107 &    0&9224  &   1&0264    \\
   80  &   1&5330     &   0&9947   &    0&9521 &    0&8639  &   0&9766    \\
   82  &   1&5078     &   0&9852   &    0&9386  &    0&8501  &   0&9658   \\
   90  &   1&3913     &   0&9500    &    0&8796 &    0&7884  &   0&9228   \\
   92  &   1&3586     &   0&9428   &    0&8640 &    0&7716  &   0&9128    \\
\hline \hline            
\end{tabular}
\end{table}
\begin{table}
\centering
\setlength{\tabcolsep}{0.5em}
\caption{The nuclear recoil effect on the $g$ factor of B-like calcium and uranium for the Coulomb and different effective screening potentials. The contributions of the zeroth ($A(\alpha Z)$, Eq.~(\ref{A_def})) and first ($B(\alpha Z)/Z$, Eq.~(\ref{B_def})) orders in the interelectronic interaction are presented together with their sum. For calcium the corresponding results for the KS and LDF potentials from Ref.~\cite{glazov:18:os} are displayed as well. 
\label{table:Breit_total}}
\begin{tabular}{ccr@{.}lr@{.}lr@{.}lr@{.}lr@{.}lr@{.}lr@{.}l}
\hline 
\hline
$Z$& Term &
     \multicolumn{2}{c}{Coul}  & 
     \multicolumn{2}{c}{CH}       &
     \multicolumn{2}{c}{PZ}       &
     \multicolumn{2}{c}{KS}      &
     \multicolumn{2}{c}{LDF}       &
     \multicolumn{2}{c}{KS \cite{glazov:18:os}}      &
     \multicolumn{2}{c}{LDF \cite{glazov:18:os}}      \\
\hline
  20  & $A$     &  $-$0&77596  &  $-$0&73398  &  $-$0&73259 &  $-$0&72669  &  $-$0&73233    &  $-$0&726622  &  $-$0&727531  \\
      & $B/Z$   &     0&09235  &     0&06012  &     0&05810 &     0&05152  &     0&05801    &     0&051451  &     0&053088  \\
      & $A+B/Z$ &  $-$0&68361  &  $-$0&67386  &  $-$0&67449 &  $-$0&67517  &  $-$0&67432   &  $-$0&675171  &  $-$0&674443  \\
\hline
  92  & $A$     &  $-$0&66745  &  $-$0&66236  &  $-$0&66157 &  $-$0&66058  &  $-$0&66206    & \multicolumn{2}{c}{} & \multicolumn{2}{c}{} \\
      & $B/Z$   &     0&01477  &     0&01025  &     0&00939 &     0&00839  &     0&00992    & \multicolumn{2}{c}{} & \multicolumn{2}{c}{} \\
      & $A+B/Z$ &  $-$0&65268  &  $-$0&65211  &  $-$0&65218 &  $-$0&65220  &  $-$0&65214    & \multicolumn{2}{c}{} & \multicolumn{2}{c}{} \\
\hline 
\hline            
\end{tabular}
\end{table}

\begin{table}
\centering
\setlength{\tabcolsep}{1.2em}
\caption{The Breit, QED, and total nuclear recoil contributions to the ground-state $g$ factor of B-like ions expressed in terms of the function $F(\alpha Z)$ defined by Eq.~(\ref{F_def}). 
\label{table:total}}
\begin{tabular}{cr@{.}lr@{.}lr@{.}l}
\hline \hline
$Z$
& \multicolumn{2}{c}{$F_{\text{Breit}}(\alpha Z)$}        
& \multicolumn{2}{c}{$F_{\text{QED}}(\alpha Z)$}          
& \multicolumn{2}{c}{$F_{\text{total}}(\alpha Z)$}        
\\
\hline
   20  &   $-$0&674(11)     &   0&0001   &    $-$0&674(11) \\
   30  &   $-$0&7076(48)     &   0&0004   &    $-$0&7072(48) \\
   40  &   $-$0&7201(27)     &   0&0009   &    $-$0&7192(27) \\
   50  &   $-$0&7231(16)     &   0&0018(1)   &    $-$0&7214(17) \\
   60  &   $-$0&7194(11)     &   0&0029(1)   &    $-$0&7165(11) \\
   70  &   $-$0&7090(8)     &   0&0045(1)   &    $-$0&7046(8) \\
   80  &   $-$0&6904(6)     &   0&0070(2)   &    $-$0&6834(6) \\
   82  &   $-$0&6854(7)     &   0&0077(2)   &    $-$0&6777(7) \\
   90  &   $-$0&6601(11)     &   0&0121(3)   &    $-$0&6480(11) \\
   92  &   $-$0&6521(13)     &   0&0138(4)   &    $-$0&6383(14) \\  
\hline \hline            
\end{tabular}
\end{table}
The total nuclear recoil correction to the $g$ factor is given by the sum of the contributions discussed above. We group them into the low-order (Breit) and higher-order (QED) parts,
\begin{align} 
\label{eq:total}
  \Delta g_\text{rec} &= \Delta g_\text{Breit} + \Delta g_\text{QED}
\,,\\
\label{eq:total-B}
  \Delta g_\text{Breit} &= \Delta g_\text{L}^\text{(1el)} + \Delta g_\text{L}^\text{(2el)} + \Delta g_\text{L}^\text{(1)}
\,,\\
\label{eq:total-Q}
  \Delta g_\text{QED} &= \Delta g_\text{H}^\text{(1el)} + \Delta g_\text{H}^\text{(2el)}
\,.
\end{align}
Here the sum of the higher-order one- and two-electron contributions is referred to as the QED part.
In Table~\ref{table:total} we present $\Delta g_\text{Breit}$, $\Delta g_\text{QED}$, and $\Delta g_\text{rec}$ in terms of the mass-ratio-independent function $F(\alpha Z)$ defined by
\begin{equation} 
\label{F_def}
  \Delta g = \frac{m}{M} F(\alpha Z)
\,.
\end{equation}
According to the definitions (\ref{eq:total})--(\ref{eq:total-Q}), the Breit part incorporates the values given in Tables~\ref{table:1el_Breit}, \ref{table:2el} (rows L), and \ref{table:1_Z_Breit}, while the QED contribution comes from Tables~\ref{table:1el_QED} and \ref{table:2el} (rows H). We have employed the values calculated with the LDF potential as the final results. There are several origins of the indicated uncertainties. First, we estimate the uncertainty due to the finite-nuclear-size effect which cannot be taken into account exactly just by considering the potential for the extended nuclei. The reason for this is that the nuclear recoil operators employed were derived for the case of a point-like nucleus. The estimates were made assuming that the relative uncertainly due to the approximate treatment of the nuclear size correction is equal to the relative value of the corresponding effect for the binding energies evaluated within the Breit approximation in Ref.~\cite{ale15}. Second, we take into account the uncertainty due to the incomplete treatment of the correlation effects which can be estimated as the spread of the values obtained with different screening potentials. However, for the two-electron contributions the uncalculated terms are probably underestimated in this way, see the related discussion in Ref.~\cite{glazov:18:os}. Therefore, for the Breit part, we calculated this uncertainty as $\Delta g_\text{L}^\text{(1)} \cdot (\Delta g_\text{L}^\text{(1)}/\Delta g_\text{L}^\text{(0)})$, where $\Delta g_\text{L}^\text{(0)}=\Delta g_\text{L}^\text{(1el)} + \Delta g_\text{L}^\text{(2el)}$ is the total Breit contribution of zeroth order in $1/Z$. Here we used the values for the pure Coulomb potential since the $1/Z$ series is no longer well defined once a screening potential is introduced. For the QED part, the uncertainty was obtained by multiplying $\Delta g_\text{QED}$ by the conservative factor $2/Z$. 

The results obtained demonstrate that the higher-order (QED) part of the nuclear recoil correction to the $g$ factor of B-like ions becomes significant for high values of $Z$. Further improvement of the theoretical accuracy requires calculations of the second and higher-order terms in the $1/Z$ expansion for the Breit part and the $1/Z$ correction to the QED part. These problems will be the subject of our future investigations.

\section{Conclusion}\label{sec:conclusion}

In the present study we conducted a numerical analysis of the nuclear recoil effect on the ground-state $g$ factor of B-like ions in the range $Z = 20$--$92$. For the first time, the higher-order QED correction was evaluated to all orders in $\alpha Z$. This contribution exceeds the total theoretical uncertainty of the nuclear recoil correction for $Z \gtrsim 60$. The low-order relativistic recoil contributions were calculated to zeroth and first orders in $1/Z$ within the Breit approximation employing the relativistic operators. The higher-order contributions in $1/Z$ were partially taken into account by means of the effective screening potential. The results of this study are indispensable for accurate theoretical predictions of the $g$ factor of highly charged B-like ions. Moreover, these results are expected to be very important for probing QED at strong-coupling regime beyond the Furry picture in the forthcoming experiments at the MPIK in Heidelberg and at the HITRAP/FAIR in Darmstadt.

\section*{Acknowledgments}
This study was supported by the Russian Science Foundation (Grant No. 17-12-01097).


\end{document}